# Inverse-Designed Phase Prediction in Digital Lasers Using Deep Learning and Transfer Learning

Yu-Che Wu, Kuo-Chih Chang, Shu-Chun Chu*

Department of physics, National Cheng kung University, No.1, University Road, Tainan City 701, Taiwan

* **Corresponding author:** Shu-Chun Chu, No. 1, University Road, Tainan City 701, Taiwan

***Tel.:*** *+886-6-27757575 ext. 65227;* ***E-mail:*** *scchu@mail.ncku.edu.tw*

**Abstract:**

Digital lasers control the laser beam by dynamically updating the phase patterns of the spatial light modulator (SLM) within the laser cavity. Due to the presence of nonlinear effects, such as mode competition and gain saturation in digital laser systems, it is often necessary to rely on specifically manually tailored approach or iteration processes to find suitable loaded phases in Digital lasers. This study proposes a model based on Conditional Generative Adversarial Networks (cGAN) and a modified U-Net architecture, with designed loss functions to inverse design the loaded phases. This study employs deep neural networks to learn an effective nonlinear relation between light field intensity and the corresponding SLM-loaded phase in simulated L-shaped digital lasers, enabling the prediction of SLM-loaded phases for both analytical and non-analytical arbitrary structured light fields. The results demonstrate superior performance on non-analytical light fields compared to the current methods in L-shaped Digital lasers. Furthermore, a transfer learning strategy is introduced, allowing knowledge obtained from one class of structured beams to be effectively reused for another, as well as to cavity-length variations. Thereby enhances generalization and improves performance under limited training data. To the best of our knowledge, this is the first deep-learning-based inverse intracavity phase design framework specifically demonstrated for digital laser systems. Providing an efficient alternative for generating structured light in other digital laser systems.

## 1. Introduction

Structured light fields—laser beams characterized by specific transverse amplitude and phase distributions—play a critical role in modern optical applications. Their importance spans a wide range of fields, including optical manipulation, biomedical imaging, laser micromachining, and free-space optical communications, all of which rely heavily on precise spatial control of the light beam. Recently, Sandile Ngcobo and colleagues proposed the concept of the "digital lasers," [1] in which a spatial light modulator (SLM) is integrated directly into the laser cavity. In this architecture, the SLM serves as one of the cavity's end mirrors, and programmable phase modulation is used to dynamically alter the intracavity modal conditions. As a result, different laser modes and beam structures can be rapidly switched in Digital lasers without the need for replacing physical optical components. However, real-world implementations are often constrained by a variety of physical limitations. These include spatially inhomogeneous gain profiles, competitive interactions among modes [2], thermally induced refractive index variations, and nonlinear or imperfect responses from optical components such as the SLM [3]. These nonlinear and non-ideal effects introduce discrepancies between the target laser field distribution and the actual lasing mode. As a result, the ability to efficiently generate and stably maintain oscillation of structured light fields has become a key challenge in digital laser system design [4].

With recent advances in SLM technology—particularly in resolution, modulation depth, phase stability, and reflective efficiency—digital lasers have demonstrated the ability to stably generate a wide variety of analytical beam modes, including Hermite-Gaussian, Laguerre-Gaussian, Ince-Gaussian(IG), Bessel-Gauss(BG), and Airy beams [1, 5]. These modes possess desirable features such as spatial symmetry, orbital angular momentum, and non-diffracting characteristics, making them particularly useful in precision machining and focused biomedical applications[4]. In addition to optical fields with analytical solutions, digital lasers can also generate arbitrary optical fields by appropriately controlling the SLM-loaded phase to provide the corresponding boundary phase distribution. Among the various approaches, two methods have been explicitly disclosed in the literature to implement such control. The first method, referred to as the "$-2\phi_L$ *phase boundary*" [6] sets the SLM loaded phase to minus twice the target field phase, thereby ensuring that the target field is converted into its counter-propagating conjugate field upon reflection from the SLM, allowing the beam to gradually evolve toward the desired shape through round-trip propagation within the cavity [6, 7]. The second method, *Gaussian-convoluted target*, regards the target field as a non-uniformly weighted fundamental Gaussian beam and designs the $-2\phi_L$ boundary phase accordingly, producing a field that is more likely to sustain stable oscillation inside the resonator [8].

However, in digital laser systems, bridging the gap between the actual and desired output fields—particularly for non-analytical arbitrary structured light fields—remains highly challenging. The former method, $-2\phi_L$ *phase boundary*, theoretically supports the stable existence of the target field within the laser cavity. However, due to propagation diffraction losses and gain competition among cavity modes in the laser system, this phase boundary may lead to either stable convergence toward a field distribution that is or is not close to the intended target, or to unstable convergence, wherein the intracavity field oscillates among multiple modes. In contrast, the latter method allows all sub-Gaussian components to propagate stably but tends to overly smooth the field boundaries, resulting in loss of fine structural details. Beyond the inherent limitations of phase-projection design methods, the nonlinear characteristics of the laser system itself and imperfections in optical components also contribute to the degradation of the laser quality. Consequently, methods that directly iteratively design the SLM projection phase based on the laser field have been developed [9]. However, this inevitably increases the complexity of the laser system and the computation time required for repeated iterative calculation of the SLM projection phase.

Achieving high-fidelity, stable output of non-intrinsic structured light fields in digital laser systems remains challenging, highlighting the need for effective phase design methods tailored to the resonator characteristics. To address this issue, we propose an inverse design approach for SLM-loaded phases for digital lasers based on cGAN model. By utilizing dataset of loaded phases and their corresponding intracavity optical field intensities, the neural network learns the effective nonlinear relationship between them. Additionally, the transfer learning strategies proposed in this study enable the trained cGAN model to successfully predict SLM-loaded phases for generating both analytical and non-analytical arbitrary optical fields in digital lasers, and transfer from one cavity length to another cavity length only needs a small amount of data. The results show that the SLM-loaded phases designed by the proposed model for generating non-analytical arbitrary structured light fields demonstrate superior performance compared with conventional methods in L-shaped digital lasers, achieving enhanced fidelity and stability in the output structured light.

## 2. Design Methods

### *2.1 Problem statement and Datasets Preparation: SLM-loaded Phases and Corresponding Converged Laser Fields*

Fig. 1 illustrates the unfolded configuration of a classical end-pumped L-shaped digital laser [5, 6, 8, 10, 11] which can be regarded as an equivalent two-mirror resonator. One of the reflective surfaces is provided by the coated crystal facet, while the other is formed by the SLM panel, where the loaded phase defines the phase boundary and thereby governs the laser modes generated in the digital laser. The objective of this study is to develop a deep learning model and learning strategy capable of directly predicting the corresponding SLM-loaded phase for generating a given target optical field in digital lasers.

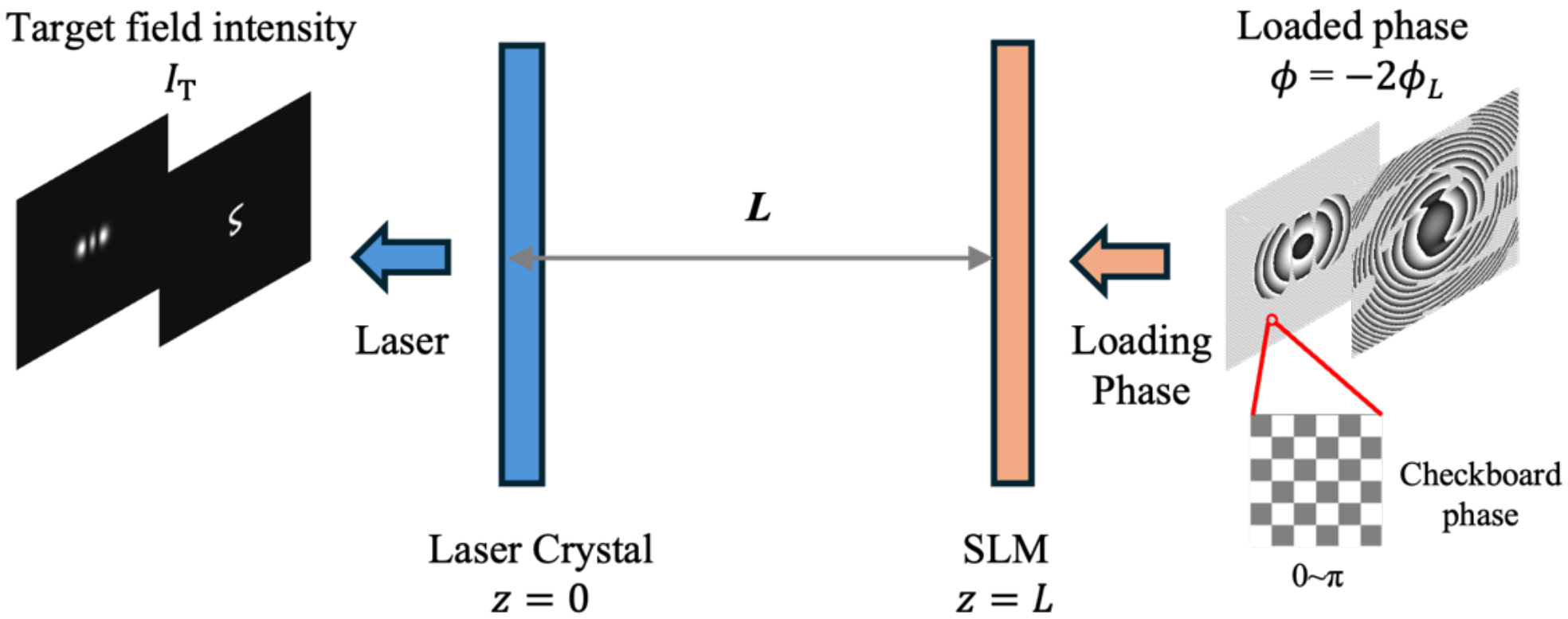

Fig. 1 An equivalent two-mirror cavity configuration for an unfolded end-pumped L-shaped digital laser.

To effectively address nonlinear effects in digital laser cavities while preserving finer spatial details of the laser fields, we generated training datasets comprising SLM-loaded phases and their corresponding laser field patterns. The SLM-loaded phases were derived using the "$-2\phi_L$ phase boundary" method. In the main energy region of the target field, the $-2\phi_L$ phase boundary ensures that the target field is converted into its counter-propagating conjugate field upon reflection from the SLM, while a 0–π checkerboard-patterned phase was applied to the low-energy regions to increase the diffraction loss of non-target components within the laser cavity. For each SLM-loaded phase, the corresponding convergent laser fields in the L-shaped digital laser cavity were calculated using the widely adopted laser cavity convergence model proposed by Dr. Endo [5-8].

These datasets were subsequently employed to train a cGAN model for the inverse design of loaded phases corresponding to specified structured fields. It is noteworthy that even though the "$-2\phi_L$ phase boundary" physically provides a suitable phase condition to support the target light field in a digital laser cavity, it is often insufficient due to the nonlinear effects of the laser. Consequently, employing the "$-2\phi_L$ phase boundary" in digital lasers frequently leads to two distinct behaviors during repeated intracavity propagation: (i) stable convergence to a field distribution (which may or may not be close to the intended target), or (ii) the absence of stable convergence, where the intracavity field oscillates among multiple patterns.

The calculation procedure for the converged laser field and simulation parameters adopted in this study were described as follows. This study adopts Endo's method, an improved Fox–Li-based simulation incorporating gain dynamics, to obtain the converged laser field. In the numerical model, Endo's gain model is incorporated to realistically describe the optical gain experienced by the light field during propagation through the laser crystal. The initial intracavity field is assumed to be a simulated spontaneous emission field with unit amplitude and a random phase distribution propagating in all directions, expressed as $U_{ini} = \exp(i2\pi R_{\#})$, where $0 \leq R_{\#} < 1$ is a computer-generated random number. Using Fresnel–Kirchhoff diffraction integration [12], the intracavity light field iteratively propagates back and forth inside the laser cavity and undergoes modulation by the cavity mirrors, spatial light modulator (SLM), and gain medium. Through repeated round trips, the converged laser field of the resonator is obtained. Fig. 1 shows the equivalent optical path diagram of the L-shaped digital laser cavity, where the effective cavity length is denoted by $L$. The simulation parameters are listed as follows. The cavity optical length $L$ was 24.8 cm, and the lasing wavelength was 1064 nm. The gain medium was a 1-mm-long Nd: $YVO_4$ crystal with an ordinary refractive index $n_o$ = 1.9573 and a circular distribution with a diameter of 1000 μm. The cavity end mirror had a diameter of 3 mm with a reflectivity of 0.99, while the optical efficiency of the SLM was set to 0.9. The SLM was discretized with a resolution of 256×256 pixels, where each pixel had a size of 12.5 μm × 12.5 μm, corresponding to a physical area of 3.2 mm × 3.2 mm. The light field intensity was also sampled with a resolution of 256×256 pixels, corresponding to a physical area of 3.2 mm×3.2 mm. For each simulation, the intracavity field was iteratively propagated for 1000 round trips to ensure sufficient convergence of the laser field.

Based on the requirements of the subsequent study, three different types of target laser field datasets were prepared: analytically defined Ince–Gaussian modes, Bessel–Gaussian beams, and non-analytical patterns derived from MNIST handwritten digits [13], which are described in detail as follows.

*a. Ince-Gaussian modes*

This study chooses IG modes as representative examples, which are complete and orthogonal analytical solutions to the Helmholtz equation in elliptic cylindrical coordinates. At the $z = 0$ plane of the laser, the IG beams in elliptic coordinate system $r = (\xi, \eta)$ can be expressed as:

$$IG^{e}_{p,m}(r, \varepsilon) = CC^{m}_{p}(i\xi, \varepsilon)C^{m}_{p}(\eta, \varepsilon)\exp\left[-\frac{r^2}{{w_0}^2}\right], \tag{1}$$

$$IG^{o}_{p,m}(r,\varepsilon) = SS^{m}_{p}(i\xi,\varepsilon)S^{m}_{p}(\eta,\varepsilon)\exp\left[-\frac{r^2}{{w_0}^2}\right], \quad (2)$$

where $w_0$ denotes the beam waist, the superscripts $e$ and $o$ indicate the even and odd modes, respectively, and $\varepsilon$ represents the ellipticity. The variables ξ and η represent the elliptic and hyperbolic coordinates in the elliptic cylindrical coordinate system, respectively, while $\varepsilon$ is the ellipticity parameter. The terms $C$ and $S$ are normalization constants, whereas $C_p^m$ and $S_p^m$ denote the even and odd Ince polynomials. According to the modal characteristics, the mode numbers $m$ corresponds to the number of hyperbolic nodal lines, while $(p-m)/2$ corresponds to the number of elliptic nodal lines. Consequently, $(p-m)$ must be zero or a positive even integer, since elliptic nodal lines cannot appear in half-numbers. For odd modes, $m \geq 1$ is required (at least one nodal line), whereas even modes allow $m=0$. These modes have been well studied and are known to be stably supported in digital laser resonators [11]. By tuning the ellipticity parameter $\varepsilon$, IG modes continuously transform between Laguerre–Gaussian (LG) profiles ($\varepsilon \to 0$) with circular symmetry and Hermite–Gaussian (HG) profiles ($\varepsilon \to \infty$) with rectangular symmetry. We further introduced $\theta$, representing the rotation angle of the transverse optical field coordinate grid, and constructed a dataset covering different mode parameters $(even/odd, p, m, \varepsilon, \theta)$. Owing to the finite effective numerical aperture of the digital laser resonator and the limited gain region size, higher-order IG modes tend to not survive in the cavity. Therefore, the ranges of the IG dataset were restricted to $p, m < 5$, with ellipticity value $\varepsilon$ equals 1000, 5, and 0.01. In total, 576 samples were generated, each containing the light fields and their corresponding loaded phases.

*b. Bessel-Gauss Beams*

Bessel-Gauss beams are analytical solutions of the paraxial wave equation in circular cylindrical coordinates, which are formed by apodizing ideal nondiffracting Bessel beams with a Gaussian envelope and are characterized by a transverse intensity profile described by the Bessel function $J_m(k_\mathrm{t} r)$. BG beams remain nearly invariant along the propagation axis $z$, and have been demonstrated to be stably generated in both conventional and digital lasers. [11, 14]. At the $z=0$ plane of the laser, the even and odd BG beams in circular cylindrical coordinates $(r, \varphi)$ can be divided into even and odd modes:

$$BG^{e}_{m}(r,\varphi;k_\mathrm{t}) = \exp(-r^2/{w_0}^2)J_m(k_\mathrm{t} r)\cos(m\varphi), \quad (3)$$

$$BG^{o}_{m}(r,\varphi;k_\mathrm{t}) = \exp(-r^2/{w_0}^2)J_m(k_\mathrm{t} r)\sin(m\varphi), \quad (4)$$

where $k_\mathrm{t}$ is the transverse wave number. The parameter $\mathrm{w}_0$ is the Gaussian term waist, and the superscripts $e$ and $o$ denote even and odd modes, respectively. The modal order $m$ corresponds to the number of topological charges or radial nodal lines. We constructed a stable BG dataset by considering the stable range of BG beams within a digital laser cavity, where stable modes are defined as those that converge in the resonator, and unstable modes are those that fail to do so. Specifically, we choose the mode order m from 1 to 5 for both even and odd modes. For each mode, the corresponding stable lasing range of $k_\mathrm{t}$ was divided into seven equal intervals, and representative samples were collected at each boundary (including the lower limit and the upper limit). A total of 80 BG data pairs is established.

*c. Non-analytical Structured Light Fields*

For non-analytical structured light fields, three types of datasets were constructed based on the MNIST handwritten digits dataset [13]: (i) Stable-only, which consisted of 50 samples of optical fields that were confirmed to stably exist within the digital laser cavity; and (ii) Stable-extended, a larger dataset of 1100 samples that included the same 50 stable fields together with an additional 1050 stable samples. The extended dataset was primarily designed to enhance the model's performance on non-analytical structured light fields and to further verify its capability in predicting the loaded phases of such fields. This dataset was also used for transfer learning across different cavity lengths. (iii) Dataset for testing different cavity lengths: the loaded phases originally designed for the cavity length of 24.8 cm, were applied directly without redesign. This dataset was constructed to evaluate the robustness of the pretrained model under cavity-length variations and to support transfer learning across different cavity lengths.

The MNIST dataset was selected as a representative example of arbitrary structured light fields based on the following considerations. First, a sufficient amount of data is required for training data-driven models. Compared with a limited number of specially designed patterns or randomly generated optical fields, MNIST provides a large number of image samples with different shapes. Second, due to the intrinsic physical limitations of the L-shaped digital laser cavity adopted in this study, the target field distributions should not contain excessively high spatial-frequency components. In particular, the finite spatial resolution of the SLM imposes a fundamental constraint on the radial phase sampling. For high-order or highly complex field distributions, the radial phase variation becomes excessively rapid, leading to insufficient phase sampling by the SLM pixels and a severe degradation of diffraction efficiency[15]. Therefore, highly complex field distributions cannot be stably supported in this configuration. The handwritten digit patterns inherently consist of relatively low-spatial-frequency strokes and are thus suitable as target images under the resolution constraint of the present L-shaped digital laser system. It is noted that generating highly complex patterns generally requires a different cavity design [9], which is beyond the capabilities of the present L-shaped configuration.

## 2.2 cGAN Model for Inverse-Designed Phase Prediction

Fig. 2 illustrates the conceptual architecture of the proposed cGAN-based inverse phase design model. In this work, the adoption of a cGAN framework is motivated by the physical requirements of intracavity phase design in digital laser systems, rather than by generic image-to-image regression performance. The adversarial loss in the cGAN explicitly promotes the preservation of high-frequency details and structurally consistent features by penalizing predicted outputs that deviate from the distribution of physically plausible samples [16]. In contrast, regression-based approaches that do not incorporate adversarial objectives—such as conventional supervised convolutional neural network models—tend to produce overly smooth phase distributions [17, 18]. Autoencoders [19, 20] in image processing tasks are primarily designed for faithful reconstruction of input data by learning a compact latent representation that minimizes reconstruction error. Variational autoencoders (VAEs) [21], on the other hand, are primarily designed to model global data distributions and continuous latent spaces and are therefore more suitable for generative diversity rather than inverse phase design.

The core framework adopted in this study is a U-Net generator combined with a PatchGAN discriminator, which follows the cGAN design introduced by Isola et al. [16], with further modifications in both architecture and loss functions. From a physical perspective, the objective of this framework is to approximate the nonlinear mapping between a target light field intensity distribution and the corresponding SLM-loaded phase. Instead of explicitly modeling the underlying cavity dynamics, the network learns this mapping directly from paired intensity–phase data generated through numerical laser simulations. The U-Net architecture, originally proposed by Ronneberger et al.[22], is characterized as a fully convolutional network with a symmetric encoder–decoder structure. This architecture enables multi-scale spatial processing of the input light field intensity distribution, allowing the network to capture both the overall intensity contour and finer spatial structures relevant to SLM-loaded phase prediction. The encoder progressively compresses spatial information from the input intensity, while the decoder outputs the corresponding SLM-loaded phase at the same spatial resolution as the input intensity. The encoder path comprises successive convolutional and pooling layers that progressively extract semantic representations from the input, while the decoder path employs upsampling operations to reconstruct the spatial resolution of the output. A distinctive feature of U-Net is its skip connections, which concatenate feature maps of the same resolution from the encoder and decoder, thereby enabling the integration of high-level semantic information with low-level texture details. However, as emphasized by Wang et al.[23], shallow skip connections may introduce semantic inconsistency, since resolution-matched feature maps do not necessarily share compatible semantic content. In the present problem, such inconsistency may lead to distortions or artifacts in the predicted SLM-loaded phase. To mitigate these issues, we strategically removed the skip connections in the first two layers of the U-Net based on empirical observations.

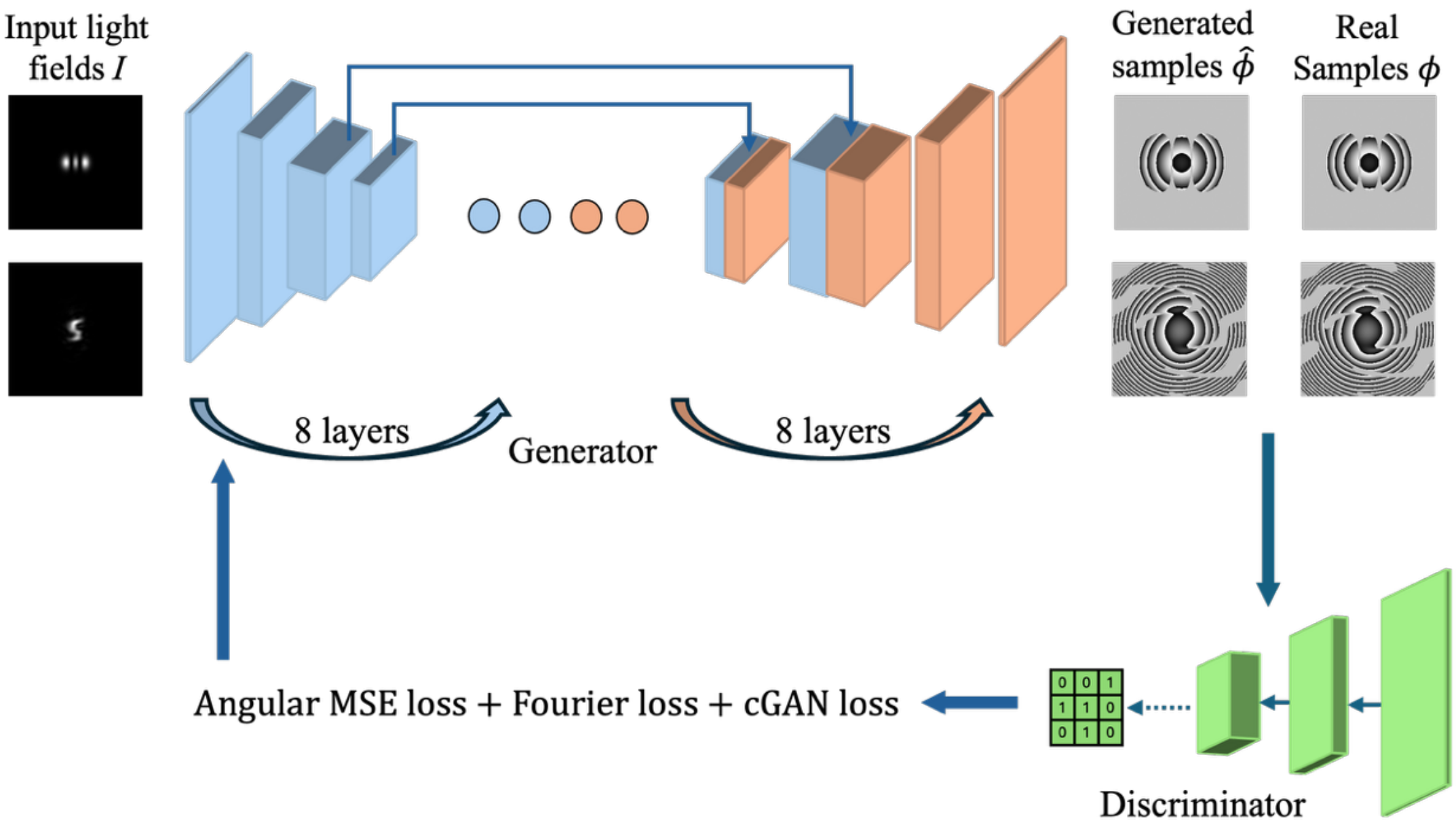

Fig. 2 Conceptual architecture of the proposed cGAN-based inverse design model. The generator (modified U-Net) maps the input light field intensity

to the corresponding SLM-loaded phase, while adversarial, MSE, and Fourier losses jointly guide training.

In contrast to the standard GAN [24], which relies solely on a random noise as input, the cGAN incorporates additional information. In this work, we do not employ a latent noise vector, as output diversity is not required for the present task. Here, the light field intensity distribution $I$ serves as the conditioning input. The generator $G$ aims to predict the corresponding SLM-loaded phase $\hat{\phi}=G(I)$, while the discriminator $D$ is trained to distinguish physically consistent real intensity–phase pairs $(I,\phi)$ which are classified as true, and inconsistent fake pairs, including generated pairs $(I,\hat{\phi})$ or mismatched combinations of real light field intensities with incorrect loaded phases, which are classified as false. Notably, the objective of cGAN is not to perform a direct pixel-wise comparison between $\hat{\phi}$ and the ground-truth $\phi$, but rather to leverage the discriminator to project the input data into a high-dimensional feature space for evaluation, which will encourage the generator to produce SLM-loaded phase distributions that are more consistent with corresponding intracavity light field intensity distributions. To further strengthen the discriminator's ability to assess both local structures and global consistency, we also adopt the PatchGAN discriminator architecture [16], which evaluates local intensity–phase image consistency across the spatial domain and outputs a confidence map that is averaged to yield the final classification score. This design promotes the generation of SLM-loaded phase patterns that are locally consistent while remaining globally and physically plausible. The objective function of the cGAN is defined as

$$\mathcal{L}_{\mathrm{cGAN}}(G,D)=\mathbb{E}_{\phi,I\sim p(\phi,I)}[\log D(\phi,I)]+\mathbb{E}_{I\sim p(I)}\left[\log\left(1-D(\hat{\phi},I)\right)\right]. \tag{5}$$

Where $\mathbb{E}[\cdot]$ denotes the expectation operator, $p(\phi,I)$ denotes the joint data distribution of paired SLM-loaded phases $\phi$ and their corresponding light field intensity $I$ generated by our L-shaped digital laser experimental process, and $\hat{\phi}=G(I)$ represents the phase predicted by the generator $G$. Following the standard GAN training strategy, the generator was trained to minimize $-\log\left(D(\hat{\phi},I)\right)$ instead of minimizing $\log\left(1-D(\hat{\phi},I)\right)$, in order to mitigate the gradient vanishing problem during the early training stage[24].

In addition to the adversarial loss, two auxiliary losses were introduced to guide the training: angular mean squared error (angular MSE) loss and Fourier loss. The angular MSE loss enforces pixel-level similarity between the generated phases $\hat{\phi}$, and the ground-truth loaded phases $\phi$. The angular MSE loss is given by:

$$\mathcal{L}_{\text{angular MSE}}=\mathbb{E}_{\phi,I\sim p(\phi,I)}[2-2cos(\hat{\phi}-\phi)]. \tag{6}$$

Since the angular MSE loss computes global differences at the pixel level, and the checkerboard pattern together with its boundaries occupy a considerable portion of the image, the angular MSE loss places greater emphasis on the checkerboard regions, their corresponding boundaries, and certain phase edges. The Fourier loss [25] compensates for high-frequency component loss that may occur due to encoder downsampling, consistent with the frequency-domain consistency principle. The Fourier loss is given by:

$$\mathcal{L}_{\mathrm{F}}=\frac{1}{2}\left(\left\| |\mathcal{F}[\hat{\phi}]|-|\mathcal{F}[\phi]| \right\|_1+\left\|\angle\mathcal{F}[\hat{\phi}]-\angle\mathcal{F}[\phi]\right\|_1\right), \tag{7}$$

Where $\|\cdot\|_1$ is sum of the absolute values, $\mathcal{F}[\cdot]$ denotes the Fourier transform, and $\angle[\mathcal{F}[\cdot]]$ and $|\mathcal{F}[\cdot]|$ denotes its phase and amplitude. Prior to applying the Fourier transform, a Hann window[26] was used to suppress spectral edge artifacts arising from finite image boundaries. This window gradually attenuates signal amplitude near the edges, leading to smoother transitions and more accurate spectral representation. Moreover, the Hann window suppresses the peripheral checkerboard, allowing the network to focus more effectively on the characteristic $-2\phi_L$ phases structural features at the phase boundary. The final objective is expressed as:

$$\mathcal{L}_{\text{total}} = \arg\min_{G}\max_{D} \mathcal{L}_{\text{cGAN}} + \lambda_1 \mathcal{L}_{\text{angular MSE}} + \lambda_2 \mathcal{L}_{\text{F}}. \tag{8}$$

The weighting parameters were set to $\lambda_1 = 20$ and $\lambda_2 = 0.7$, the batch size was set to 10.

Overall, the deep learning model adopted in this study followed the original cGAN framework [16] with several modifications. The input to the network is a scalar light field intensity distribution. In the generator, the U-Net architecture was modified by removing the skip connections in the first two layers to reduce semantic inconsistency. The PatchGAN discriminator was simplified to three convolutional layers with 32, 64, and 128 filters, respectively. The normalization layers were all instance normalization[27], rather than batch normalization[28], to improve stability during training. Optimization was performed using the Adam solver[29], with learning rates of 0.0005 for the generator and 0.0001 for the discriminator, and momentum parameters set to $\beta_1 = 0.9$ and $\beta_2 = 0.99$.

### *2.3 Transfer Learning Strategy*

Transfer learning enables knowledge acquired in a source task to be reused in a related target domain, which is particularly valuable when labeled data are scarce and limited [30]. Based on this concept, this study proposes a transfer learning strategy, as illustrated in Fig. 3, which shows the workflow of the proposed framework. As an initial validation, IG modes were used to demonstrate that the cGAN model can learn correspondence between structured light fields and their associated loaded phases, and be the source task. The network is first pretrained on the IG dataset in Section 2.1 (a) for 200 epochs, using learning rates of 0.0005 for the generator and 0.0001 for the discriminator. Transfer learning is implemented using a fine-tuning all layers. All network parameters are copied from the pretrained model and further trained on the target task without freezing any layers. The optimizer is reinitialized before fine-tuning, and the pretrained network weights are used as the initialization for subsequent training[31]. This strategy is adopted because freezing intermediate layers in a U-Net architecture with skip connections may introduce discontinuities between frozen and trainable layers, which frozen weight was observed in our experiments to degrade performance.

Subsequently, transfer learning is investigated in three representative scenarios. First, analytical BG modes were used to demonstrate that reliable prediction can be achieved with only limited data through transfer learning. The pretrained IG weights are transferred to a new network, and the model is fine-tuned on the BG dataset in Section 2.1 (b) for 200 epochs using learning rates of 0.0003 for the generator and 0.0001 for the discriminator. Reduced learning rates are adopted because IG and BG modes exhibit similar spatial field distributions. Second, the MNIST handwritten digits dataset [13] was adopted as a representative non-analytical target field to demonstrate how transfer learning can improve prediction accuracy when the training datasets include poorly designed phases that fail to achieve stable laser output. For the Stable-only MNIST dataset described in Section 2.1(c), the model was fine-tuned for 200 epochs using learning rates of 0.005 and 0.001 for the generator and discriminator, respectively. For the larger Stable-extended MNIST dataset, using learning rates of 0.0005 and 0.0001 for the generator and discriminator, respectively. Third, to investigate the effectiveness and advantages of transfer learning when the system undergoes variations or when discrepancies exist between the training system and the target system (e.g., experimental errors such as optical misalignment, imperfect components, and measurement inaccuracies), transfer learning across different cavity lengths is investigated. As described in Section 2.1(c), the network is first pretrained on the Stable-extended MNIST dataset generated at a cavity length of 24.8 cm for 200 epochs, using learning rates of 0.0005 and 0.0001 for the generator and the discriminator. The pretrained weights are then transferred, and the model is fine-tuned on the target cavity-length datasets for 200 epochs, using smaller learning rates of 0.00005 for the generator and 0.00001 for the discriminator.

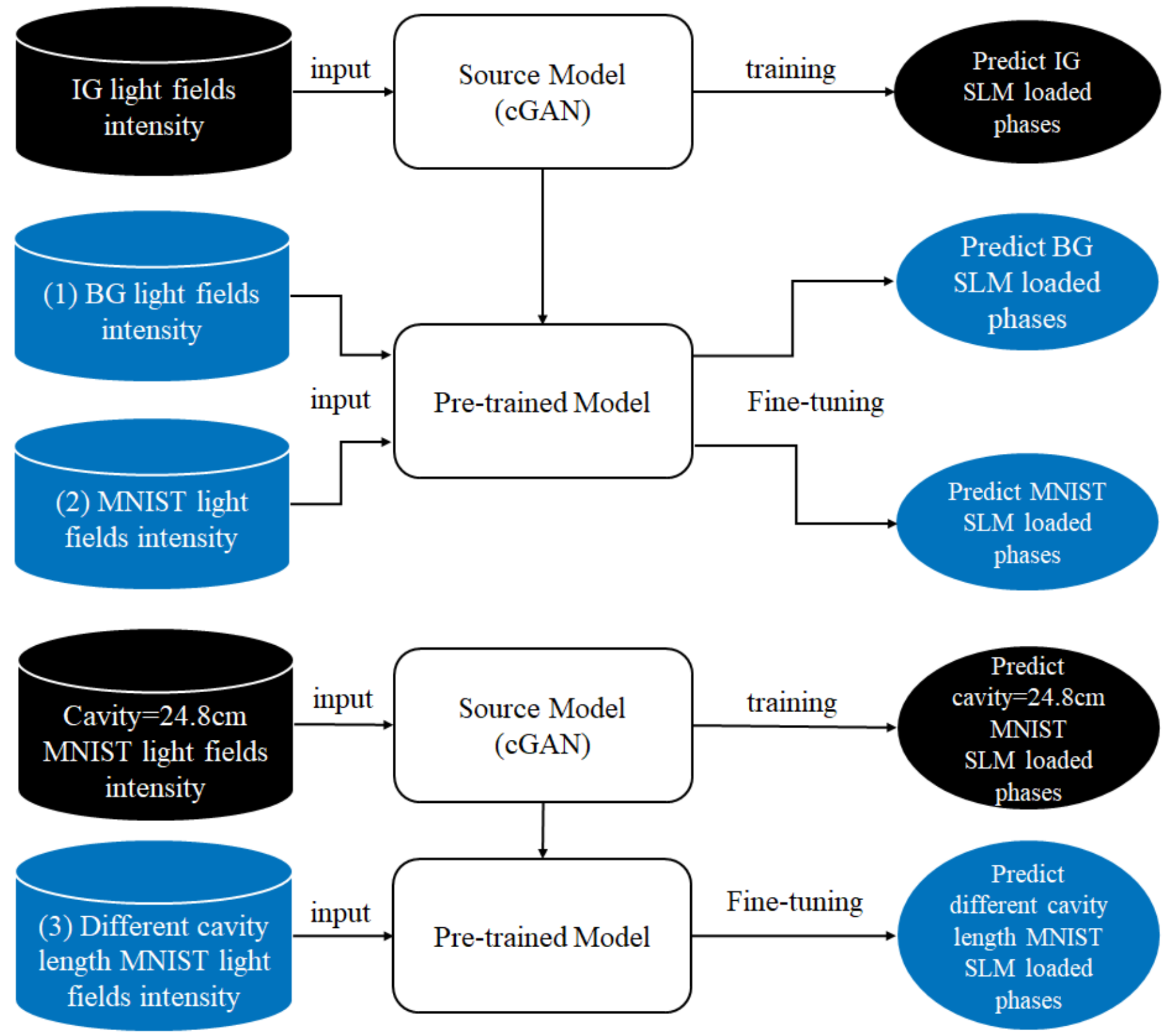

Fig. 3 Workflow of the proposed transfer learning framework for inverse-designed phase prediction in digital lasers. A cGAN is first trained on IG modes to serve as a source model and then fine-tuned on BG modes or MNIST light fields. Next, a cGAN is trained on cavities with 24.8 cm and then fine-tuned to different cavity lengths to predict the corresponding SLM-loaded phases

## 3. Results and Discussion

In this section, Section 3.1 first presents the results of the cGAN model learning, and its successful prediction of the digital laser projection phases required to generate IG modes, and then demonstrates how the pretrained cGAN model, through transfer learning, can more efficiently generalize to predict the projection phases necessary for generating another class of analytical light fields, namely BG beams. Section 3.2 presents the outcomes of applying the cGAN model in combination with a transfer learning approach, highlighting its effectiveness in accurately predicting and generating non-analytical structured light in L-shaped digital laser systems. Section 3.3 further examines the degradation of model performance under variations in cavity length and evaluates the effectiveness of transfer learning to different cavity lengths.

In the following, the performance of the cGAN model, either after standard training or transfer learning, is evaluated by comparing the reconstructed light fields intensity for each test sample with the corresponding target fields intensity $I_{\mathrm{T}}$. Two loaded phases are prepared: the loaded phases $\phi$ obtained via previous $-2\phi_L$ phase boundary, and the predicted loaded phases $\hat{\phi}$ generated by the cGAN model. Both loaded phases are then applied to the L-shaped laser resonator, and the resulting convergent light intensities, $I$ (from $\phi$) and $\hat{I}$ (from $\hat{\phi}$), are computed through the simulated intracavity wave iteratively propagation modal. The similarity of $I$ and $\hat{I}$ with the target field intensity $I_{\mathrm{T}}$ are then evaluated separately to verify the physical plausibility and reconstruction fidelity of the predicted loaded phases. This study uses the uncentered correlation coefficient and the normalized root mean square error (NRMSE) as the criteria function, which is denoted uncentered correlation as correlation and defined as

$$\text{Correlation} = \frac{\iint [I_{\mathrm{re}}(x,y) \cdot I_{\mathrm{T}}(x,y)]\, dx\, dy}{\sqrt{\iint I_{\mathrm{re}}(x,y)^2\, dx\, dy} \cdot \sqrt{\iint I_{\mathrm{T}}(x,y)^2\, dx\, dy}}. \tag{9}$$

An NRMSE is defined as

$$\text{NRMSE} = \sqrt{\frac{\iint |I_{\mathrm{re}}(x,y) - I_{\mathrm{T}}(x,y)|^2\, dx\, dy}{\iint |I_{\mathrm{T}}(x,y)|^2\, dx\, dy}}, \tag{10}$$

where $I_{\mathrm{re}}$ denotes the reconstructed light field. Depending on the specific case, $I_{\mathrm{re}}$ may correspond either to $I$ (from $\phi$) and $\hat{I}$ (from $\hat{\phi}$).

### *3.1 Validation of the cGAN Model Prediction and Transfer Learning Strategy*

This section presents the validation of the proposed cGAN model and its transfer learning capability. First, the effectiveness of the cGAN model is demonstrated through learning and predicting SLM-loaded phases of analytically defined IG modes. Subsequently, the pretrained model is adapted via transfer learning to predict SLM-loaded phases of another analytically defined optical field, specifically BG beams, illustrating the model's generalization across different laser patterns.

The model was first trained on the IG dataset introduced in Section 2.1(a) for 200 epochs. To evaluate whether the trained model is capable of predicting the loaded phase to reconstruct target IG modes, 100 unseen IG mode configurations were selected as the test set. These test samples span a range of configurations $(even/odd, p, m, \epsilon, \theta)$, providing a diverse distribution to evaluate the model's generalization performance on unseen data. The correlation and NRMSE values of the 100 testing IG samples are plotted as histograms in Fig. 4 (a) and (b) comparing the cGAN (trained on IG modes) and the $-2\phi_L$ phase boundary method relative to the target fields $I_{\mathrm{T}}$. The reconstructed and light intensity obtained via cGAN and $-2\phi_L$ phase boundary method of the parameter sets (e,3,1,5,3), (e,4,0,1000,177), (e,4,4,5,148), (o,4,4,0.01,148) ,and (o,4,2,0.01,85) are illustrated in the insets of the figure, corresponding to (i)–(v), respectively.

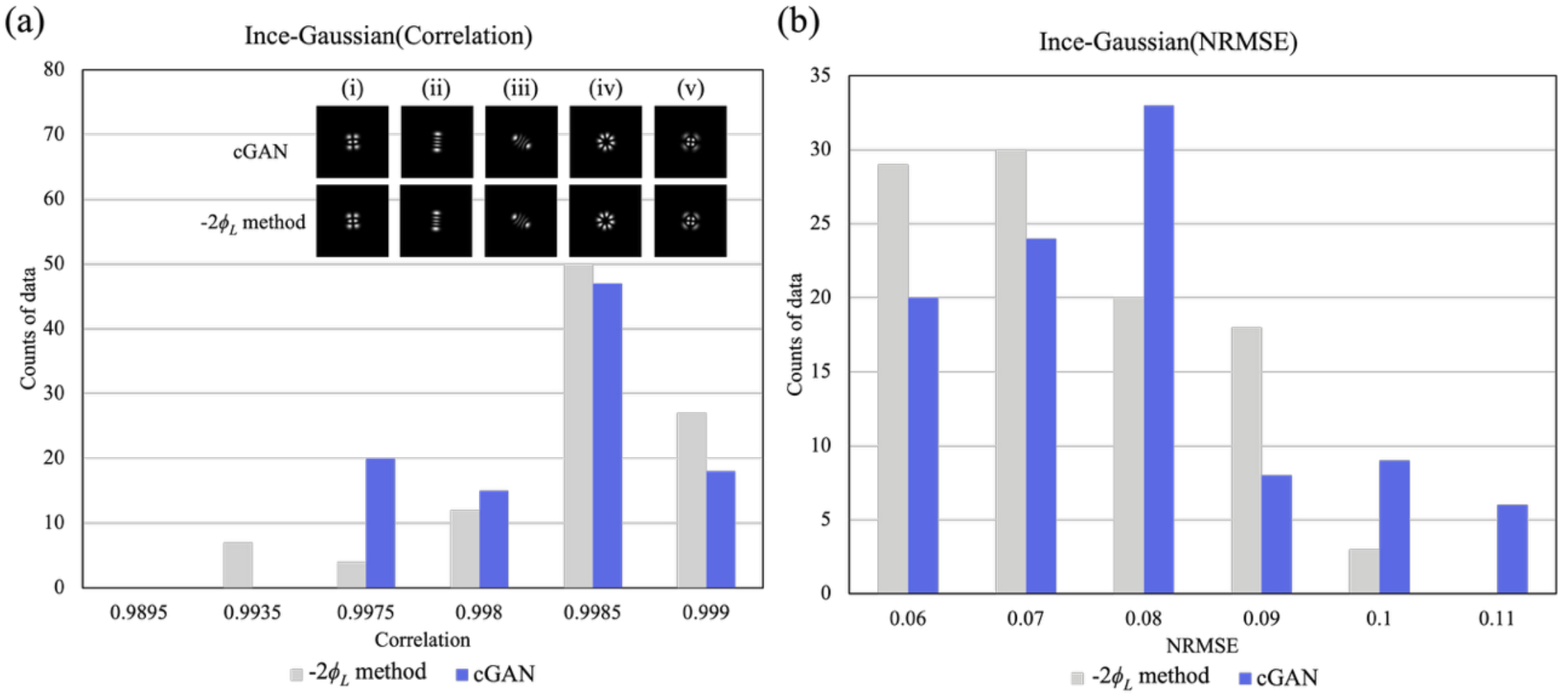

Fig. 4 Statistical distribution of (a) correlation values and (b) NRMSE values for reconstructed IG light fields obtained using the cGAN (trained on IG modes) and the $-2\phi_L$ phase boundary method relative to target intensity, evaluated on 100 unseen IG configurations. Subfigures (i)–(v) show representative reconstructed fields intensity $\hat{I}$ compared with reconstructed light fields $I$ obtained via $-2\phi_L$ phase boundary method for parameter sets (e,3,1,5,3), (e,4,0,1000,177), (e,4,4,5,148), (o,4,4,0.01,148), and (o,4,2,0.01,85).

All test samples achieved correlation scores above 0.9895 and NRMSE scores below 0.11, confirming that the model preserves strong spatial fidelity in predicting loaded phases for reconstructing IG light fields. These results indicate that the reconstructed light fields $\hat{I}$ generated by the model exhibit a high degree of spatial consistency with the target fields intensity $I_{\mathrm{T}}$ validating that the model has learned the correspondence between structured fields and their loaded phases. Moreover, the reconstruction quality is comparable to light fields intensity $I$ obtained via $-2\phi_L$ phase boundary method. These findings show the model's ability to handle the learned structured IG light fields.

To further evaluate transferability under limited data conditions, we used a stable BG dataset consisting of 80 samples to train two models for 200 epochs. One model was initialized from scratch, while the other was pretrained on IG modes. We then compared their predictive capability for loaded phases of BG modes. The test set consisted of 210 unseen BG configurations $(even/odd, m, k_{\mathrm{t}})$. Fig. 5 (a) and (b) show the statistical distribution of the correlation values and NRMSE values of reconstructed fields $\hat{I}$ relative to the target fields intensity $I_{\mathrm{T}}$, comparing the BG-only training approach with the IG→BG transfer learning on the 210 unseen BG modes. The representative parameter sets (e,1,5250), (e,2,5750), (o,3,8500), (o,4,10750), and (e,5,2750) are illustrated in the subfigure, corresponding to (i)–(v), respectively.

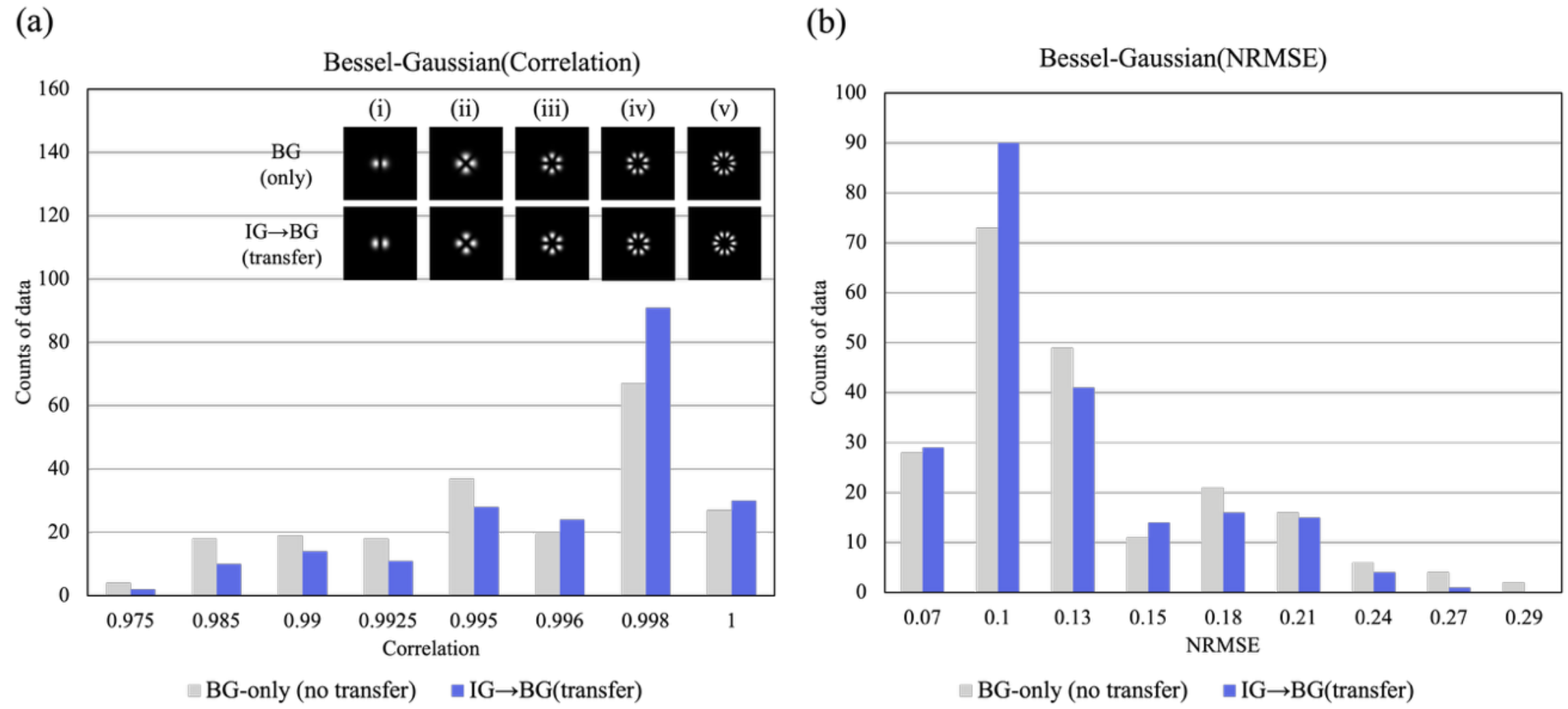

Fig. 5 Statistical distribution of (a) correlation values and (b) NRMSE values of reconstructed BG modes on 210 unseen configurations relative to the target fields intensity $I_{\mathrm{T}}$, comparing the cGAN model trained on the stable BG dataset with cGAN transfer from IG mode to BG mode. Subfigures (i)–(v) show representative reconstructed fields intensity $\hat{I}$ comparing the BG-only training approach with the IG→BG transfer learning for parameter sets (e,1,5250), (e,2,5750), (o,3,8500), (o,4,10750), and (e,5,2750).

For the model trained solely on BG modes, it exhibits lower correlation distribution and higher NRMSE distribution, indicating that the model struggles to learn the characteristic features of BG beams from a limited dataset. In contrast, the transfer learning from IG to BG modes shifts the correlation distribution toward higher correlation values and lower NRMSE values. Although IG and BG modes belong to different modal families, they exhibit underlying structural correlations. Pretraining on IG modes thus provides physically meaningful priors, embedding essential spatial structures and resonator characteristics relevant to BG beams. Leveraging IG modes as the source task positions the pretrained model on the manifold already proximal to the BG modes solution space in latent space. With this advantageous initialization, only a few additional BG samples are needed to refine the representation and achieve accurate BG predictions. As a result, the model can capture more robust cavity features and effectively adapt to the non-diffracting BG regime. The results demonstrate that the proposed transfer learning approach achieves improved accuracy and generalization compared with directly training the model from scratch.

### *3.2 Transfer learning for generating non-analytical structured light*

This section presents the application of a transfer learning strategy to extend the capability of the trained cGAN model, enabling it to successfully predict the SLM-loaded phase required to generate arbitrary non-analytical optical fields in digital lasers, thereby overcoming the nonlinear effects inherent in the system and efficiently achieving stable oscillation of structured light fields.

To evaluate the model's ability to transfer to non-analytical structured light fields, we trained two models on the Stable dataset for 200 epochs as mentioned in Section 2.3: one from scratch and the other pretrained on IG modes using learning rates of 0.005 and 0.001 for the generator and discriminator, respectively. The test set consisted of 200 unseen non-analytical light fields, and the reconstruction fidelity was assessed using correlation and NRMSE. Fig. 6 shows the statistical distribution of the correlation and NRMSE values of reconstructed fields $\hat{I}$ relative to the target fields intensity $I_{\mathrm{T}}$, comparing the Stable-only training approach, and IG→Stable transfer learning on the 200 unseen data.

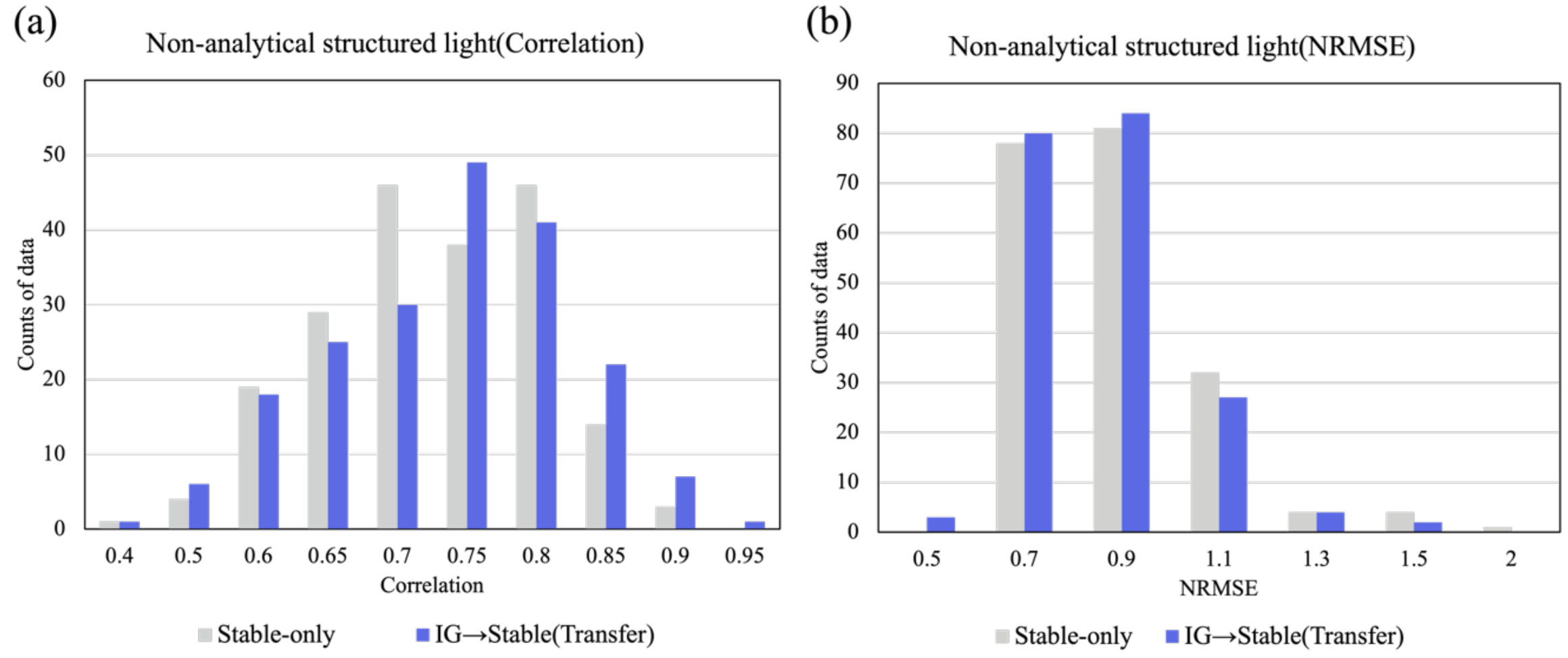

Fig. 6 Statistical distribution of (a) correlation values and (b) NRMSE values of reconstructed non-analytical light fields $\hat{I}$ relative to the target field intensities $I_{\mathrm{T}}$, comparing the Stable-only training approach with the IG→Stable transfer learning strategy on 200 unseen samples.

As shown in Fig. 6 (a) and (b), the model trained from scratch exhibits lower correlation values and higher NRMSE values. With transfer learning, the model outperforms training solely on the Stable-only dataset, suggesting that features learned from analytical IG modes can be effectively leveraged to improve prediction performance on non-analytical structured light fields under limited data conditions.

Fig. 6 reveals that transfer learning, even with only 50 samples, achieves reliable phase prediction for generating non-analytical light fields. To further improve the prediction accuracy of the model, we expanded the dataset to a Stable-Extended in Section 2.1(c) and trained the pretrained model on IG modes for 200 epochs, using the same test dataset as in the previous implementation. A subsequent comparison with the conventional $-2\phi_L$ phase boundary method. Fig. 7 shows a comparison of several reconstructed fields intensity $\hat{I}$ of cGAN, $-2\phi_L$ phase boundary method and the target fields intensity $I_{\mathrm{T}}$. The first row shows the reconstructed fields intensity $\hat{I}$ from cGAN, the second row presents the reconstructed fields intensity $I$ from $-2\phi_L$ phase boundary method. The third row displays the target fields intensity $I_{\mathrm{T}}$, the fourth and fifth row shows the corresponding SLM-loaded phase generated by cGAN and $-2\phi_L$ phase boundary, respectively. Fig. 8 (a) shows the correlation values and (b) shows the NRMSE values of reconstructed fields $\hat{I}$ relative to the target fields intensity $I_{\mathrm{T}}$, comparing the cGAN training approach with the $-2\phi_L$ phase boundary method on the same 200 unseen dataset.

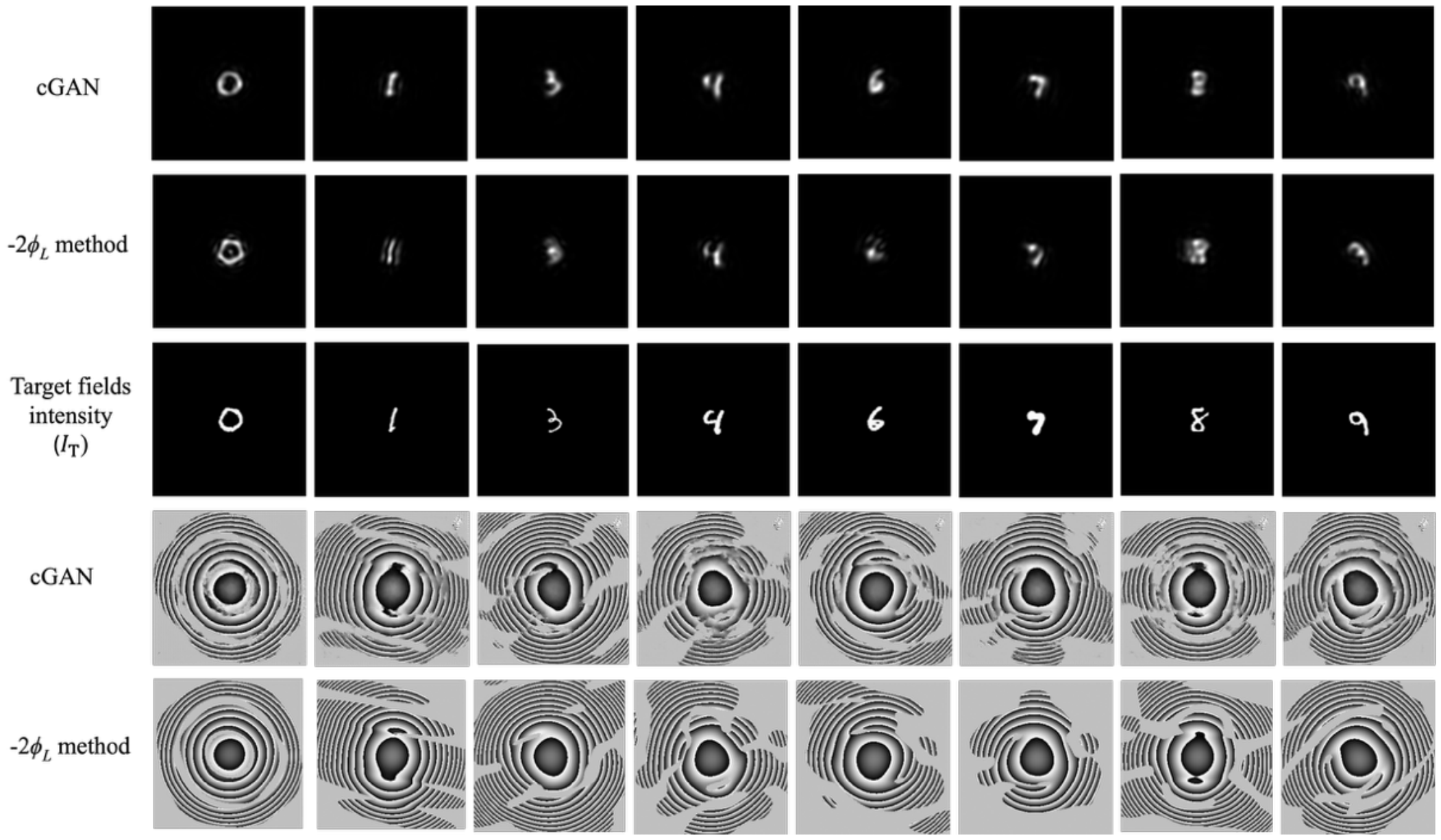

Fig. 7 Some demonstration of reconstructed fields $\hat{I}$ of cGAN,$-2\phi_L$ phase boundary method and the target fields intensity $I_{\mathrm{T}}$. The associated SLM-loaded phase designed by the cGAN and the $-2\phi_L$method are also shown.

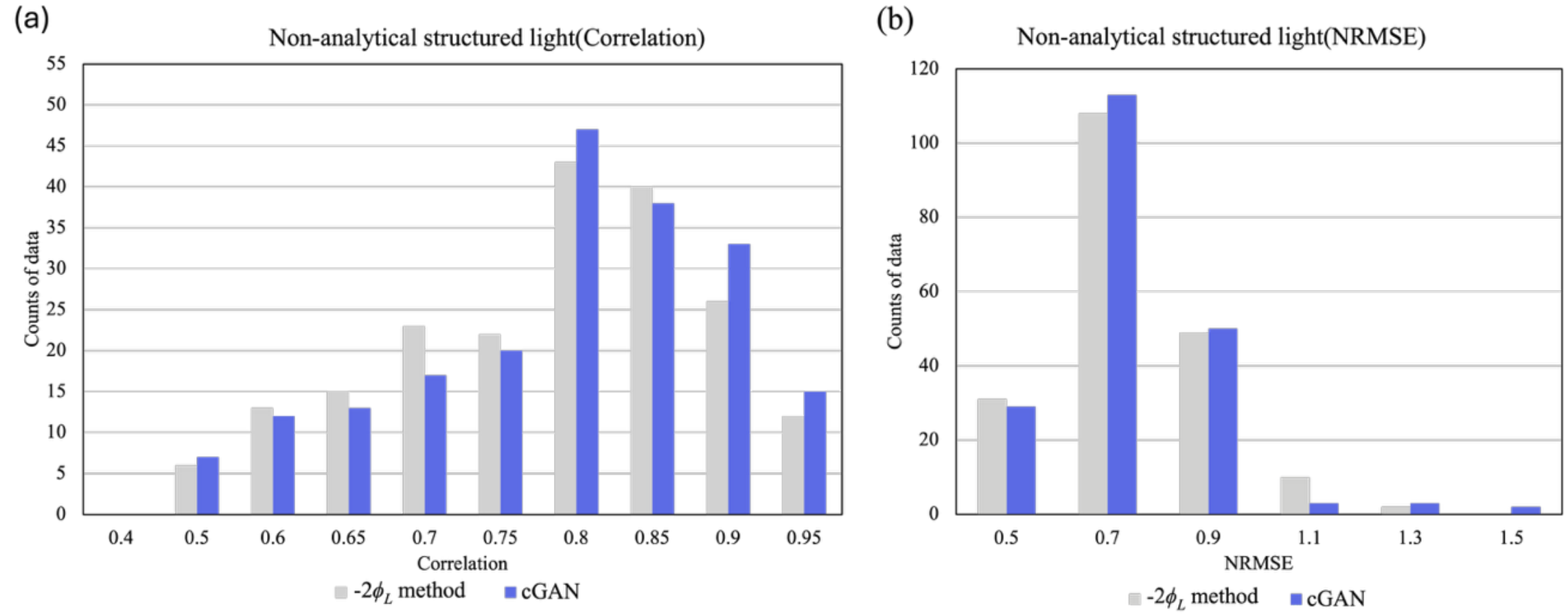

Fig. 8 Statistical distribution of (a) correlation values and (b) NRMSE values of reconstructed MNIST handwritten digits light fields on 200 unseen data relatives to the target fields intensity $I_{\mathrm{T}}$, comparing the cGAN model transfer from IG dataset to Stable-extended dataset with the $-2\phi_L$ phase boundary method

As shown in Fig. 7, for certain target fields, the conventional method fails to converge to a stable intracavity solution or even to generate the corresponding light field. In contrast, the cGAN trained with transfer learning on the Stable-Extended dataset can successfully map these target fields to convergent SLM-loaded phases, thereby producing physically plausible and stable outputs. In the fourth and fifth row, it can be observed that the cGAN indeed produces phase boundaries that slightly differ from those obtained with the conventional $-2\phi_L$ approach. Since $-2\phi_L$ phase boundary method is derived from an idealized computational phase boundary, when applied to non-analytical target fields, the nonlinear effects within the cavity or constraints on the cavity length often prevent the loaded phases from converging to a stable light field or from generating the corresponding light field. This result confirms that the framework approach proposed in this study effectively reduces the difference between the actual and desired output fields of non-analytical, arbitrarily structured light fields in digital lasers. The statistical results in Fig. 8 (a) and (b) further validate this trend. The conventional method exhibits lower correlation values and higher NRMSE values, reflecting reduced reliability for non-analytical light fields. In contrast, the cGAN achieves higher correlations and lower NRMSE distribution, indicating improvements in both accuracy and reliability. These results confirm that, with sufficient training data, the cGAN can effectively capture the effective nonlinear relation between light field intensity and SLM-loaded phase within the cavity, thereby surpassing conventional methods and enabling stable reconstructions even in cases where the conventional approach fails to converge. Overall, the well-trained cGAN model can consistently predict loaded phases that enable the intra-cavity light field to stably converge to a distribution close to the intended target field intensity $I_{\mathrm{T}}$ , with its predicted phase achieving superior performance in generating non-analytical structured light fields compared to conventional methods.

Considering overall computational efficiency, the limited resolution and numerical aperture of the resonator adopted in this study inherently impose some restrictions on the performance of the digital laser, particularly because the restricted angular spectrum within the cavity further limits the spatial resolution of the generated light fields. It is therefore expected that increasing the numerical aperture of the digital laser cavity would enable the generation of light fields with higher clarity and improved spatial resolution. While the present simulations adopt a simplified resonator model, the framework is inherently scalable. A promising direction is to first train the model within numerical simulations and then transfer the learned representations to real-world digital laser systems, thereby compensating for experimental nonlinearities and imperfections. Such a simulation-to-experiment transfer pathway would enhance the practical feasibility of AI-driven inverse design.

### *3.3 Transfer learning for generating non-analytical structured light under cavity-length variations*

To investigate whether the proposed neural network can be transferred to practical digital laser systems in the presence of experimental imperfections—such as collimation errors and component misalignment—we use variations in cavity length as a controlled surrogate for such system-level deviations. Specifically, as described in Section 2.3, the cGAN model trained on a cavity length of 24.8 cm was directly applied to digital laser cavities with different lengths without any additional fine-tuning, in order to evaluate its generalization capability. Fig. 9 (a) summarizes the resulting performance degradation under cavity-length mismatch, with the NRMSE shown on the right vertical axis in reversed order to visually align its decreasing trend with the correlation. The model, trained at a cavity length of 24.8 cm, was evaluated using 200 unseen MNIST non-analytical target fields. For each test case, the reconstructed intensity $\hat{I}$ was compared with the target intensity $I_T$ using the correlation and NRMSE, and the metrics were averaged over the dataset. The plotted results correspond to cavity lengths of 23.4, 24.1, 24.8, 25.5, and 26.2 cm, which correspond to approximately ±3% and ±6% deviations from the nominal cavity length. As the cavity length progressively deviates from the training condition, a monotonic degradation is observed, indicating that a neural network trained for a specific L-shaped cavity configuration cannot maintain optimal performance under cavity-length mismatch. This behavior suggests that unavoidable experimental deviations in practical systems can reduce prediction accuracy, motivating the need for transfer learning.

We therefore consider a practical transfer learning scenario in which a neural network trained on a well-characterized source system—such as a numerical simulation or a well-aligned digital laser—is adapted to a target cavity configuration using only a small number of stable samples. As described in Section 2.3, the network was first pretrained at a cavity length of 24.8 cm using the Stable-extended dataset, and we selected the 25.5 cm configuration as a representative target system for demonstrating transfer learning. Only 20, 50, and 100 stable samples were used for fine-tuning at the target cavity length. Fig. 9 (b) compares the performance of the transfer learning approach with models trained from scratch under the same data availability, with the NRMSE shown on the right vertical axis also in reversed order. The reconstructed intensity $\hat{I}$ was evaluated against the target intensity $I_T$ using the correlation and NRMSE, and the results were averaged over the dataset. For all three training-set sizes, the transfer learning approach consistently achieves higher correlation and lower NRMSE than training from scratch. Notably, even with as few as 20 training samples, the transferred model already outperforms the model trained from scratch. These results demonstrate that transfer learning enables efficient adaptation to cavity-length variations using only a small amount of target-cavity length data.

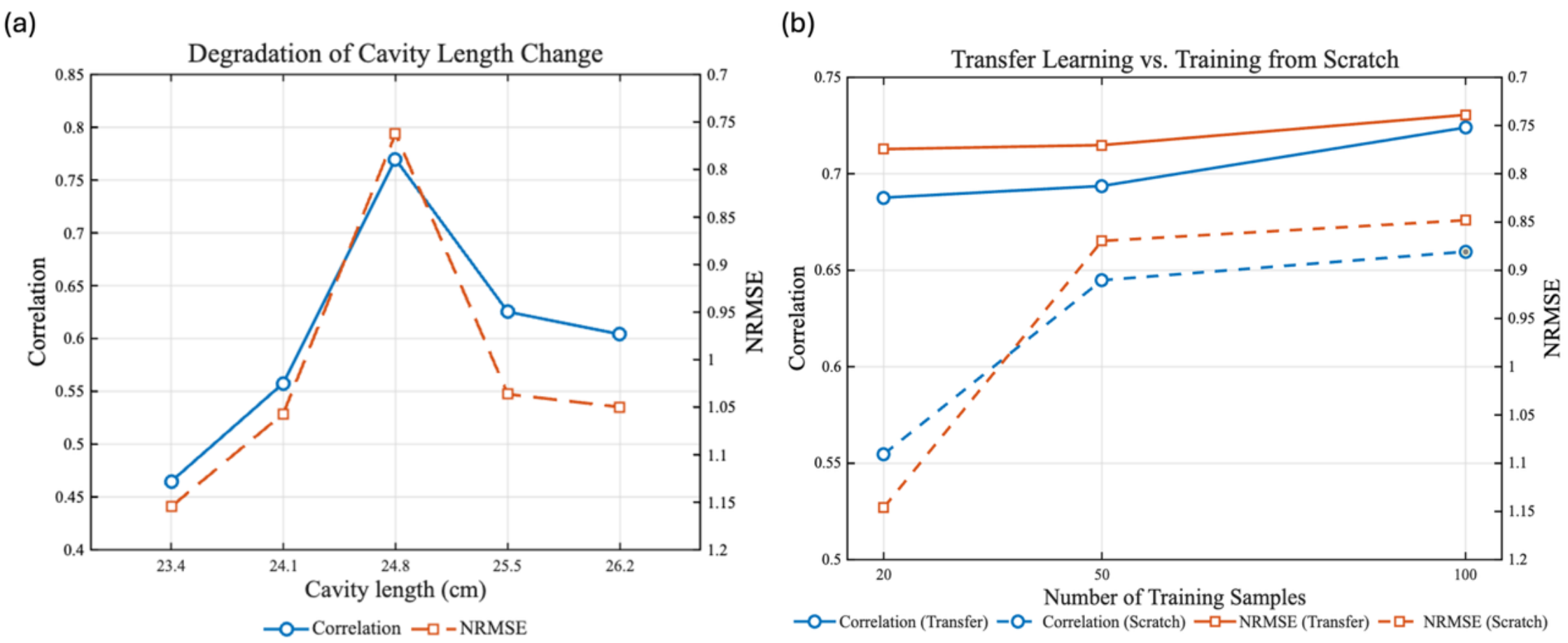

Fig. 9 Effect of cavity-length variation on model and transfer learning efficiency. (a) Degradation of reconstruction fidelity as the cavity length deviates from the training condition. (b) Comparison between transfer learning and training from scratch under limited training samples. The right vertical axis corresponds to the NRMSE, which is plotted in reversed order for visual consistency (lower NRMSE indicates better reconstruction performance).

While the present simulations adopt a simplified resonator model, the proposed framework is inherently scalable. A promising direction is to first train the model within numerical simulations and subsequently transfer the learned representations to real-world digital laser systems, thereby compensating for experimental nonlinearities, misalignments, and other experimental inherent properties. As demonstrated by the cavity-length variation numerical experiments, transfer learning provides an effective way for adapting the model to cavity length deviations using only a limited amount of target-domain data. In practical experimental environments, additional factors such as background noise and measurement fluctuations are unavoidable. These effects could be addressed by incorporating a dedicated denoising module, such as an autoencoder [32, 33], to map noisy measured field intensities to their corresponding noise-suppressed representations before phase prediction [34]. Such a simulation-to-experiment transfer pathway would further enhance the practical feasibility of data-driven inverse design in digital laser systems.

We now turn to a discussion of the limitations, underlying implications, and broader applicability of the proposed framework. The proposed framework constitutes a data-driven nonlinear fitting approach that aims to approximate the SLM-loaded phase distributions capable of generating desired target light fields. This characteristic complements the shortcomings of traditional theoretical methods in handling non-analytical problems and offers a new pathway for the design and control of digital lasers. Relying solely on paired intensity–phase data, the method is independent of explicit cavity parameters such as resonator length, gain region size, or pump position. It should be noted, however, that the proposed model does not learn the underlying laser cavity physics. Similar to other data-driven approaches, the network effectively learns an empirical correction to the $-2\phi_L$ phase boundary method within the distribution spanned by the training data. As a result, its performance and generalization are inherently influenced by data availability, representativeness, and common limitations associated with purely data-driven models [35]. While the proposed framework is formulated in a data-driven framework based on paired intensity–loaded phase samples and could, in principle, be extended to other cavity geometries, such extensions would require dedicated simulations or experiments in future work.

A natural question arises as to why the training dataset is generated using the $-2\phi_L$ phase boundary method combined with iterative intracavity propagation, while the proposed model is observed to outperform the $-2\phi_L$ method in certain cases. One possible explanation is that the $-2\phi_L$ approach provides a theoretical phase boundary derived from idealized assumptions and does not explicitly account for the nonlinear effects, such as mode competition and diffraction loss in the resonator. In contrast, the training dataset is constructed from stable light field intensities obtained after iterative round-trip propagation, and therefore implicitly encodes information about the cavity's behavior. Moreover, the loss function and architecture may guide the network toward SLM-loaded phases that are more likely to remain stable in the cavity. As a result, although the network does not learn the complete laser physics, it may learn to map the input intensity to an effective phase representation, thereby improving performance over the original $-2\phi_L$ method within the sampled regime.

Notably, the proposed transfer learning strategy is not restricted to a specific class of structured light fields; the choice of source and target domains can be flexibly adapted according to the intended application. While this study demonstrates the framework within digital laser systems, the methodology is not confined to this domain. A wide range of end-to-end image-to-image control problems, where the mapping between structured inputs and outputs is nonlinear and difficult to model explicitly, could potentially benefit from this type of data-driven and transfer learning framework.

## 4. Conclusion

In this study, we proposed a data-driven inverse phase design framework for digital laser systems, integrating cGAN, a modified U-Net architecture, and a composite loss function design. The framework directly predicts the SLM-loaded phase from the intended light field intensity. For non-analytical structured light fields, such as handwritten digit patterns, the proposed cGAN demonstrates stable convergence toward intensity distributions close to the intended targets, whereas the conventional $-2\phi_L$ phase boundary method often fails to converge. Trained on analytically solvable IG modes, the model achieves high-fidelity reconstructions and preserves spatial consistency with the target fields intensity. With transfer learning, the pretrained model can be efficiently adapted with limited additional data to non-diffracting BG beams, outperforming models trained from scratch. The proposed transfer learning strategy is also shown to enable efficient adaptation across different cavity lengths, surpassing training from scratch. These results confirm the capability of the framework to provide a practical alternative to manual trial-and-error or iterative phase design processes in Digital lasers. Overall, this study highlights the potential of deep learning and transfer learning for digital lasers. Looking ahead, integrating physics-informed neural networks (PINNs) [36], the framework could evolve into a more physically consistent model, paving the way toward robust and interpretable control of structured light fields in practical systems.

**Funding.** National Science and Technology Council (NSTC 114-2112-M-006-018-).

**Disclosures.** The authors declare no conflicts of interest.

**Data availability.** Data underlying the results presented in this paper are not publicly available at this time but may be obtained from the authors upon reasonable request.